# Assessing the Readiness of Greece for Autonomous Vehicle Technologies

Chrysostomos Mylonas[1], Charis Chalkiadakis[2], Alexandros Dolianitis[3], Dimitris Tzanis[4], Evangelos Mitsakis[5]

[1]Research Associate, Centre for Research and Technology Hellas, Hellenic Institute of Transport, Thermi, Thessaloniki
E-mail: chmylonas@certh.gr

[2]Research Associate, Centre for Research and Technology Hellas, Hellenic Institute of Transport, Thermi, Thessaloniki
E-mail: charcal@certh.gr

[3]Research Associate, Centre for Research and Technology Hellas, Hellenic Institute of Transport, Thermi, Thessaloniki
E-mail: adolianitis@certh.gr

[4]Research Associate, Centre for Research and Technology Hellas, Hellenic Institute of Transport, Thermi, Thessaloniki
E-mail: dtzanis@certh.gr

[5]PhD, Senior Researcher, Centre for Research and Technology Hellas, Hellenic Institute of Transport, Thermi, Thessaloniki
E-mail: emit@certh.gr

## Abstract

Despite the debate regarding the timeframe and rate of penetration of Autonomous Vehicles, their potential benefits and implications have been widely recognized. Therefore, assessing the readiness of individual countries to adopt such technologies and adapt to their introduction is of particular importance. This paper aims to enrich our understanding of EU readiness regarding the introduction of autonomous vehicle technologies by assessing the case of Greece. Thus, through a literature review, the criteria upon which such an assessment should be based are established and analyzed. Subsequently, the case of Greece is assessed based on those criteria by finding relevant sources that support and justify any assessment. Regardless of the outcome concerning the readiness of Greece, such an assessment should help identify areas in which focus should be given in order to ensure a smoother transition to such technologies. This contribution is expected to assist policy makers in Greece.

## Περίληψη

Παρά την έλλειψη ομοφωνίας σχετικά με τη χρονική στιγμή και το βαθμό που τα Αυτόνομα Οχήματα θα διεισδύσουν στην αγορά, τα δυνητικά τους οφέλη και οι επιπτώσεις τους έχουν αναγνωριστεί σε μεγάλο βαθμό στη σχετική βιβλιογραφία. Κατά συνέπεια, η αξιολόγηση της ετοιμότητας μεμονωμένων κρατών να υιοθετήσουν και να προσαρμοστούν στην έλευση αυτών των οχημάτων κρίνεται ιδιαίτερα σημαντική. Η παρούσα εργασία στοχεύει να εμπλουτίσει την τρέχουσα αντίληψη σχετικά με την ετοιμότητα της Ευρωπαϊκής Ένωσης να εισάγει Αυτόνομα Οχήματα στα συστήματα μεταφορών της, μέσω της αξιολόγησης της ετοιμότητας της Ελλάδας. Συγκεκριμένα, μετά την επισκόπηση της σχετικής βιβλιογραφίας, τα κριτήρια στα οποία μια τέτοια αξιολόγηση οφείλει να βασιστεί αναγνωρίζονται και αναλύονται. Κατόπιν, η περίπτωση της Ελλάδας αξιολογείται επί τη βάσει αυτών των κριτηρίων με τη συνδρομή πηγών και βάσεων δεδομένων από τη βιβλιογραφία. Ασχέτως των αποτελεσμάτων σχετικά με την περίπτωση της Ελλάδας, μια τέτοια ανάλυση αναμένεται να συντελέσει στην αναγνώριση συγκεκριμένων περιοχών στις οποίες πρέπει να επιμείνουν οι αρμόδιοι εθνικοί φορείς ώστε να διευκολύνουν τη μετάβαση σε αυτή την ενδιαφέρουσα τεχνολογική εξέλιξη.

## *1. Introduction*

Autonomous Vehicles (AVs) appear to be the cornerstone for a decisive reshaping of the road transport sector. This is evident from the rising level of interest both within academia and the industry. Such an interest may be translated to various research topics including technical details of AV technologies and their policy implications. While the former is mainly concerned with issues such as the applications of Artificial Intelligence and Deep Learning techniques on the design and coding of such vehicles, the latter encompasses issues such as the legal and social implications of AVs, their implications for transport planning and mobility patterns, as well as how cities should adapt to the paradigm shift resulting from the emergence of AVs. Another interesting topic that falls into the second of the aforementioned categories is the assessment of the readiness of each country to introduce such vehicles into their transport networks. This paper aims to elaborate further on this topic by: a) reviewing the available literature, b) shaping a set of representative criteria, c) weighing their criticality by conducting an Analytic Hierarchy Process (AHP) analysis amongst the authors, and d) assessing the readiness of Greece. This type of analysis is considered of particular importance when taking into account that regardless of the final outcomes with respect to the case of Greece, several areas could be identified in which policy makers should focus in order to enable the smooth transition to AV technologies.

The rest of this paper is organized as follows. Section 2 includes a brief review of the literature dealing with the topic at hand. Section 3 builds on this review by distinguishing the criteria that a nation-wide assessment of the readiness for AVs should be based on. Section 4 presents the adopted procedure for and the results of the weighting of said criteria. Section 5 assesses the case of Greece based on the adopted criteria, while Section 6 provides the reader with the derived conclusions.

## *2. Literature Review*

The topic of assessing the readiness of a country for AVs appears to be of increasing interest for the global research community. However, given that the overarching topic of AVs is still relatively new, the number of relevant studies that can be found in the literature is somehow limited.

A first indicative example may be found in annual study by KPMG that was first conducted in 2018. This study has recently been updated in 2019 and is expected to be updated again in the following years. Within this study 25 countries around the world have been ranked based on their apparent readiness to introduce AV technologies. This assessment examines four overarching criteria categories, namely those of "Policy and Legislation", "Technology and Innovation", "Infrastructure", and "Consumer Acceptance" (KPMG, 2019).

The category of "Policy and Legislation" examines issues such as AV regulations, the existence of an AV department within the governmental structure, the governmental change readiness, the effectiveness of law making bodies, the efficiency of the legal system in changing regulations, the number of government funded AV pilots, and finally an assessment of the data sharing environment (KPMG, 2019).

The category of "Technology and Innovation" examines issues such as the existence of industry partnerships, the number of AV technology firms that have their headquarters located within the country under examination, the number of AV related patents, the level of investment on AV related firms, the availability of the latest technologies, the capacity for innovation, and finally the market share of electric cars (KPMG, 2019).

The category of "Infrastructure" examines issues such as the coverage of electric vehicle charging stations, the score of global connectivity index for infrastructure as calculated by GSMA, the coverage of the 4G network, the quality of roads, the score on the Logistic Performance Index (LPI) for Infrastructure, and finally the score on the change readiness of technology infrastructure. Logistics performance is included in such an analysis taking into consideration that freight is expected to be one of the first users of AVs.

Finally, the category of "Consumer Acceptance" takes into account the results of consumer surveys on AV acceptance, the percentage of population living in test areas, an assessment of the change readiness for technology use, the technology readiness as calculated by the World Economic Forum, and finally the market penetration of ride sharing (KPMG, 2019).

Table 1 presents the criteria categorization adopted by KPMG (2019) and provides further details regarding the nature and the meaning of the individual criteria.

*Table 1: AV readiness criteria proposed by KPMG (2019)*

| Category | Criterion | Description/Details/Data sources |
|---|---|---|
| Policy and Legislation | AV regulations, government-funded AV pilots, and AV-focused agency | Regarding the sub-criterion "AV regulations", it is of interest whether a country has drafted regulations that support AV deployment and place restrictions on when, where, and how testing of AVs may occur |
| | | A similar approach applies to the sub-criterion "government-funded AV pilots" focusing mainly to their amount |
| | | Regarding the sub-criterion "AV-focused agency", it is of interest to examine whether a country spread the responsibility for AVs across a large number of government entities; the countries that have established a dedicated agency for AV technology and innovation are considered as advantageous |
| | Government readiness for change | This criterion relies on a complex index calculated by KPMG for various countries that includes assessments of regulation, government strategic planning, and the rule of law |
| | Effectiveness of legislative process & efficiency of the legal system in challenging regulations | This criterion relies both on the World Economic Forum's Networked Readiness Index and judgements by business executives in each country; effectiveness is meant as the ease of passing the required regulations for the deployment of AVs |
| | Data-sharing environment | This criterion relies on the World Wide Web Foundation's Open data barometer; the countries that share data in an open manner are considered advantageous |
| Technology and Innovation | Industry partnerships | Partnerships between vehicle makers and technology suppliers; based on: a review of news coverage (incl. local and global media), research conducted by consulting firms, and the content of blogs maintained by AV industry experts |
| | AV technology firm headquarters | Based on: a list of AV-related technology companies composed by those published by Vision System Intelligence and Comet Labs enhanced with data from the from Crunchbase Pro on AV companies founded since the 2018 report |
| | AV-related patents | Based on data from PatSeer analysis on all AV-related patents and patent applications |
| | Industry investments in AV | Based on data from Crunchbase Pro; it is focused on the countries of investing organizations and not on the countries where investments are made |
| | Availability of the latest technology & capacity for innovation | This criterion relies on both the World Economic Forum's Networked readiness index and the responses from business executives |
| Infrastructure | Market share of electric cars | Mainly based on data from the International Energy Agency's Global EV outlook 201; it is also based on country-specific data sources |
| | Density of EV charging stations | Mainly based on data from the International Energy Agency's Global EV outlook; it is also based on data from the US Bureau of Transportation Statistics and country-specific data sources |
| | Quality of mobile internet | As assessed by the GSM Association acting as a representative of mobile network operators; the assessment takes into account the availability of high-performance mobile internet network coverage and the number of servers and network bandwidth |
| | 4G coverage | This criterion reflects the importance of AVs having wide access to mobile data networks; it is based on data from researcher OpenSignal |

| | Quality of roads | Based on the World Economic Forum's Global competiveness report |
|---|---|---|
| | Logistics infrastructure | This criterion reflects the quality of roads specifically for logistics, using the World Bank's Logistic performance index 2018 |
| | Technology infrastructure change readiness | Based on KPMG International's Change readiness index |
| Consumer acceptance | Consumer opinions of AVs | Responds to questions such as "*What is your general opinion of autonomous vehicles?*" |
| | Population living in test areas | The rationale of this criterion is based on the assumption that as the public does see the AVs on the roads, the more familiar they will be to them and the more likely will be to trust and use them when they become available; countries in which a large amount of people live in cities where AVs are being tested are considered advantageous |
| | Civil society technology use | Based on the relevant sub-indicators of KPMG International's Change readiness index |
| | Consumer adoption of technology | Based on the World Economic Forum's Global competiveness report dealing with the availability of latest technologies, mobile broadband subscriptions, internet access and internet bandwidth |
| | Online ride-hailing market penetration | Based on data from Statista concerning the percentage of people in each country that have utilized a ride-hailing service |

The second identified study (Johnson, 2017) investigates mostly the infrastructure aspects of AV readiness without focusing on a specific country. It examines infrastructure not solely from a carriageway perspective but rather the entirety of infrastructure assets that accompany road transportation. Characteristically, it looks into communication infrastructure, roadside structures, the roads themselves, geotechnical features, and drainage. It looks into the aforementioned categories under various driving scenarios, namely platooning, advanced braking and collision avoidance systems, valet parking assistance. Table 1 includes a lower level presentation of the aforementioned criteria groupings. This qualitative analysis may be meant more as a guideline for approaching the issue rather than an assessment tool.

***Table 2:*** *AV readiness criteria from an infrastructure perspective as proposed by Johnson (2017)*

| Category | Criterion | Description/ Details |
|---|---|---|
| Communications | Roadside communication | Roadside devices that supplement vehicle-based devices, sensors, and vehicle-to-vehicle communications (incl. beacons located at strategic positions that are e.g. in position to replace traffic signals, provide vehicle position information and serve other functions as well) |
| | Fibre optic networks | Fibre optic cables connecting roadside communication devices (used instead of radio or wireless means) |
| | Construction plans | Timely availability and up-to-dateness of plans affecting the road network and being essential e.g. for mapping activities |
| | Multiple traffic signals | Adequate availability of multiple traffic signals with the aim of tackling poor visibility issues and avoiding the disruption of AVs relevant perceptual abilities |
| | Clarity of road markings, signals and signage | Quality and maintenance frequency of road markings, signals, and signage |
| | Level of standardization of signals and signage | Ensure interoperability of relevant AVs functionalities and avoid the disruption of AVs perceptual abilities |
| | Handling tolls | Automation of toll stations in order for them e.g. to be able to cope with platoons and to detect the presence of a (responsible) human driver in a passing vehicle |
| Structures | Parking facilities | Dedicated parking spaces for rental AVs positioned near certain areas (e.g. transit hubs and train stations) |
| | Fueling and power distribution | Adequate availability of (mostly electric) power supply infrastructure especially within parking facilities |
| | Segregated infrastructure | Separated road infrastructure for AVs and non-AVs (e.g. bridges and underpasses) as well as among different road users |
| | Street lighting | Adequate illumination and density of street lights to support AV visibility |
| | Roundabouts | Availability of roundabouts in certain areas (especially in areas of AVs and non-AVs coexistence) |
| Roads | Maintenance | Maintenance frequency of road markings, signs, and signals, as well as of road surface condition |

| | Autonomy-enabled roads | Existence of roads, dedicated lanes on existing lanes, and areas designated as AV-only |
|---|---|---|
| | Road geometry | Existence of narrower streets and tighter corner radii with the objective of minimizing road space and costs especially in new housing estates |
| Geotechnical features | - | Reduction of road gradients in order to harmonize speeds taking into account the introduction of platooning and convoys; Clear distinction between carriageways, sidewalks and road verges in order not to disrupt sensory systems that depend on line of sight (incl. control of vegetation to ensure adequate visibility) |
| Drainage | - | Adequate drainage capacity of road networks with the aim of avoiding to test AV competencies against surface water and flooding |

The third study (Kimley-Horn, 2016) is an indicative example of a local analysis of the actions needed to enhance a single state's readiness for Connected and Autonomous Vehicles (CAVs). This study revolves around the identification and clarification of seven key initiatives on the road to CAV readiness. These initiatives, all of which have to do with laws and policies, infrastructure, or business aspects of the issue, are the following:

- Group Structure and Organization
- Political Leadership Engagement
- Changes to Laws and Motor Vehicle Codes
- Long-Range Transportation Plans
- Mobility and Access Improvements
- Pilot Projects and Research

On a similar note, a study by Fagnant and Kockelman (2015) examines how to prepare any nation rather than a specific state for autonomous vehicles. Characteristically, this study identifies barriers to the deployment of AV technologies, in order concrete policy recommendations to address said barriers. The identified barriers focus around vehicle costs, AV certification, litigation, liability, and perception, security, privacy, and finally research gaps. As a solution, the authors suggest the expansion of governmental funding for AV research, the development of official guidelines for AV certification, and finally the formulation of appropriate standards to address liability, security, and data privacy.

Regarding AV research funding, Fagnant and Kockelman (2015) highlight the relatively low understanding of how AV will affect the transport system as a whole. They highlight key missing links in AV research such as the incorporation of market penetration scenarios in planning efforts. Furthermore, they note the role that governmental funding should play in this endeavor. Regarding, the guidelines for the certification of AVs, the authors highlight the need for a unified approach through a single and unilaterally accepted framework rather than individual states or countries addressing the issue on their own. In addition, they propose that addressing the issues of liability, security, and data privacy, through the creation of appropriate standards, needs to strike a balance between assigning responsibilities to manufactures and putting pressure on AV products.

Finally, Henderson (2018) poses special emphasis to the need for drafting appropriate frameworks and models capable of judging the readiness of AVs to operate on public streets. Particularly, taking into account that AVs should be at least as good - and as safe - as average human drivers, the author proposes a standard similar to that used for assessing the readiness of humans to drive on public streets. This is expected to facilitate the avoidance of conflicts between manufacturers and governments in issues such as liability sharing. The proposed standard includes the following criteria for the passengers or drivers of AVs in order to associate their competencies with the various levels of automation: a) minimum age, b) adequate vision, c) a specified amount of safe driver under the direct supervision of an adult with a valid driver's license, d) passing a test of traffic laws and signs, and e) passing a driving test by a certified driving instructor/evaluator.

## *3. Determination of Assessment Criteria*

This section aims to determine the criteria that an assessment of the nation-wide readiness for AVs should be based on. As it may easily become evident from the literature review, the broadest and the most direct framework on the topic at hand is that of KPMG (2019). Particularly, this framework captures the majority of the dimensions of what is considered as the transport geography (Figure 1) without adding excessive complexity by avoiding to include criteria that are less relevant with the readiness of a country for AVs. In addition, said framework is more precisely oriented to an assessment exercise considering that it utilizes tangible metrics associated with the content of various databases of internationally recognized organizations, such as the World Economic Forum.

Consequently, it was considered appropriate to adopt the aforementioned framework and to enrich it, where needed, with findings from the rest of the reviewed literature. The proposed set of criteria is also based on the following principles:

- Avoid overlaps between different criteria
- Avoid adopting vague criteria, which are not clearly associated with the topic at hand and may be addressed from more than one and, potentially, conflicting perspectives
- Avoid excessively increasing the total number of criteria by merging those that are similar in nature and adopting proposed categorizations
- Avoid using criteria that are purely intended to make recommendations rather than assessing the readiness for AVs

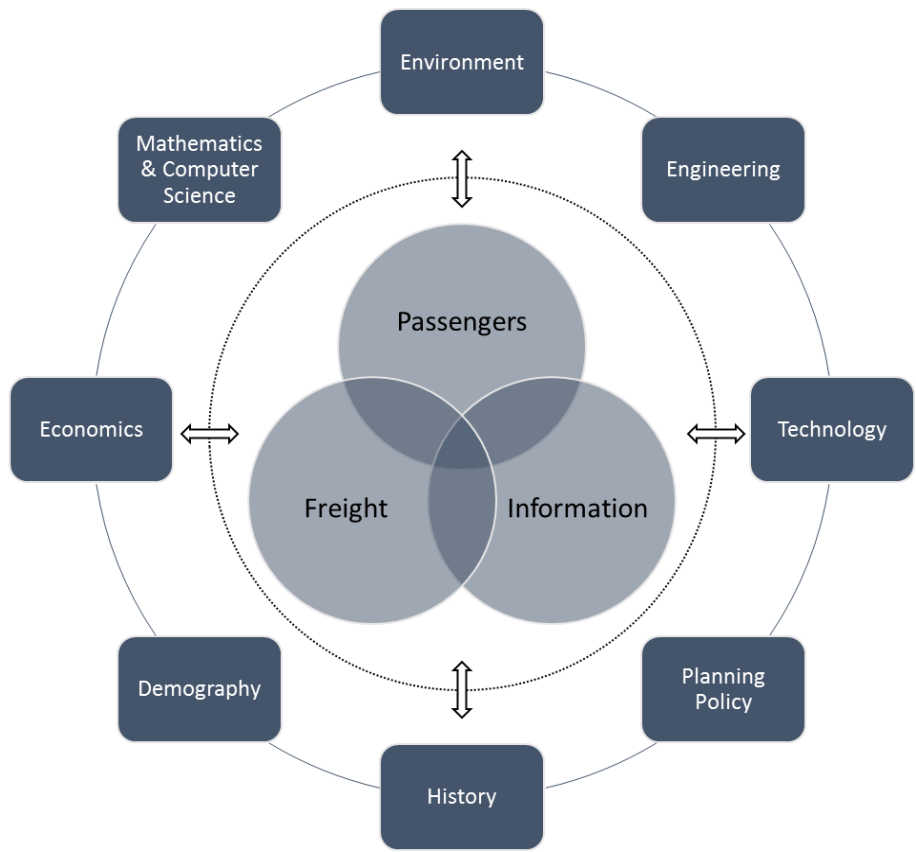

***Figure 1:*** *Transport geography [Adapted from: Rodrigue, 2017]*

Regarding the first principle, it should be noted that a possible criterion dealing with the revision of long-term transport plans that may emerge from the initiatives proposed by Kimley-Horn (2016) overlaps to a degree with the criterion "government readiness for change" as proposed by KPMG (2019). This is attributed to the fact that the latter takes into account the strategic planning capacity of a government and, therefore, its ability to adjust and adapt its long-term transport plans. In addition, the content of the criteria proposed by Johnson (2017) that deal with road maintenance, street lighting, and the drainage of the road network are to a great extent covered by the criterion "road quality" as that is proposed by KPMG (2019). Furthermore, the criterion "geotechnical features" is also covered to an extent by the aforementioned criterion proposed by KPMG (2019), while the aspects that are not covered more closely resemble suggestions and their exclusion is covered by the fourth principle. Finally, a criterion entitled "Standards and models for liability, security, and data privacy" that may emerge from the proposals of Fagnant and Kockelman (2015) may be considered as part of the effective of the legislative process criterion since even if such legislation and standards are not yet in place an effective legislative body could easily and timely propose and adopt them.

Regarding the second principle, it should be noted that the criterion "roundabouts" proposed by Johnson (2017) is not selected since it is assumed that the connectivity features and the perceptual abilities of AVs are in position to reduce the need for such traditional traffic management tools. Moreover, the criterion "fibre optic networks" proposed by Johnson (2017) is also not selected taking into account that such networks are quite expensive and may be applicable only for certain urban environments.

Regarding the third principle, it should be noted that the criteria "parking facilities" and "fueling and power distribution" proposed by Johnson (2017) and the criterion "density of EV charging stations" proposed by KPMG (2019) can be merged into a single criterion designated as "suitability of roadside structures". Moreover, with the aim of further reducing the total number of criteria, the criterion "AV technology firm headquarters" proposed by KPMG (2019) and a possible criterion that may derive from the initiatives proposed by Kimley-Horn (2016) dealing with the research institutes that are engaged with AVs are merged into a single criterion designated as "number of non-governmental AV actors". Furthermore, two criteria entitled "Education programs for AVs" and "Conducted Expos and Workshops for AVs" that emerge from the initiatives proposed by Kimley-Horn (2016) are merged into a single criterion entitled "organized promotion of AV technologies". Finally, the criteria "Quality of mobile internet" and "4G coverage" both of which have been adopted by KPMG (2019) are considered as different ways of viewing similar information and are, therefore, merged into a single criterion entitled "Quality of mobile internet".

Regarding the fourth principle, it should be noted that as already stated parts of the criterion "geotechnical features" may be considered as recommendations and may not be used as assessment criteria since no infrastructure outside testbeds currently exists that was built with them in mind. Similarly, the criteria "autonomy-enabled roads" and "road geometry" proposed by Johnson (2017) are also excluded. Furthermore, the criteria "multiple traffic signals" and "segregated infrastructure" proposed by Johnson (2017) are considered excessive and likely redundant as AV technologies evolve and are better able to perceive their environment. Finally, the criterion "determination of required competencies for using AVs" which may emerge from the proposals of Henderson (2018) is also excluded taking into account that one of the benefits of AVs is their possible use from a wider previously excluded audience (e.g. under-aged users, elderly users, and visually impaired users).

Based on the reviewed literature, the authors of this paper also propose one additional criterion. This criterion relates to consumer acceptance and measures the "average vehicle value". This is based on the argument (Nunes and Hernandez, 2019) that the cost of self-driving cars is one the most significant barriers for their adoption by the wider public.

Table 3 includes a complete list of the criteria that are to be used for the assessment of the readiness of a country towards AVs as well as their categorization.

*Table 3: List and categorization of adopted assessment criteria*

| ID | Criterion | Category |
|---|---|---|
| PL1 | AV regulations | Policy and Legislation |
| PL2 | Government-funded AV pilots | Policy and Legislation |

| PL3 | AV-focused agencies | Policy and Legislation |
| PL4 | Government readiness for change | Policy and Legislation |
| PL5 | Effectiveness of legislative process & efficiency of the legal system in challenging regulations | Policy and Legislation |
| PL6 | Data sharing environment | Policy and Legislation |
| PL7 | Organized promotion of AV technologies | Policy and Legislation |
| TI1 | Industry partnerships | Technology and Innovation |
| TI2 | Number of non-governmental AV actors | Technology and Innovation |
| TI3 | AV-related patents | Technology and Innovation |
| TI4 | Industry investments in AV technologies | Technology and Innovation |
| TI5 | Availability of the latest technology & capacity for innovation | Technology and Innovation |
| TI6 | Market share of electric cars | Technology and Innovation |
| I1 | Suitability of roadside structures | Infrastructure |
| I2 | Quality of mobile internet | Infrastructure |
| I3 | Quality of roads | Infrastructure |
| I4 | Logistics infrastructure | Infrastructure |
| I5 | Technology infrastructure change readiness | Infrastructure |
| I6 | Adequacy of roadside communication | Infrastructure |
| I7 | Shareability of construction plans | Infrastructure |
| I8 | Clarity and level of standardization of road markings, signals and signage | Infrastructure |
| CA1 | Consumer opinions regarding AVs | Consumer Acceptance |
| CA2 | Population living in test areas | Consumer Acceptance |
| CA3 | Civil society technology use | Consumer Acceptance |
| CA4 | Consumer adoption of technology | Consumer Acceptance |
| CA5 | Online ride-hailing market penetration | Consumer Acceptance |
| CA6 | Average vehicle value | Consumer Acceptance |

## *4. Weighting of Adopted Criteria*

As already stated in the introductory section of this paper, the criticality of the adopted criteria is assessed based on the AHP method. AHP constitutes a well-known structured technique for addressing complex decision-making problems that was originally developed by Thomas L. Saaty in the 1970s (Saaty, 1987). Since then it has been utilized extensively in group decision making in a wide range of research fields (Saaty, 1987). A typical use of the method across literature is the determination of the weights of the criteria to be used in a decision-making exercise. This process revolves around a pairwise comparison of the various criteria according to a numerical scale ranging from 1 to 9, the respective formulation of the pairwise comparison matrix, the normalization of said matrix, and finally the computation of the criteria weight vector.

In this particular case, the participants are limited to the five members of the author team in an effort to come to a rational consensus on the weights that are to be applied to each criteria and criteria category. Characteristically, for the purpose of this paper a tiered approach was adopted (Figure 2) meaning that five AHP sessions were conducted, one for each criteria category and the fifth one assessing the weights of the categories themselves. To this end, the variation and tool developed by Prof. Goepel (2018) was utilized. The weight of each criterion included on Table 3 is therefore the product of its own weight times the weight of the category that belongs to.

Table 4 includes the aggregated results of the weighting process and the consensus levels for each criterion and criteria category AHP session.

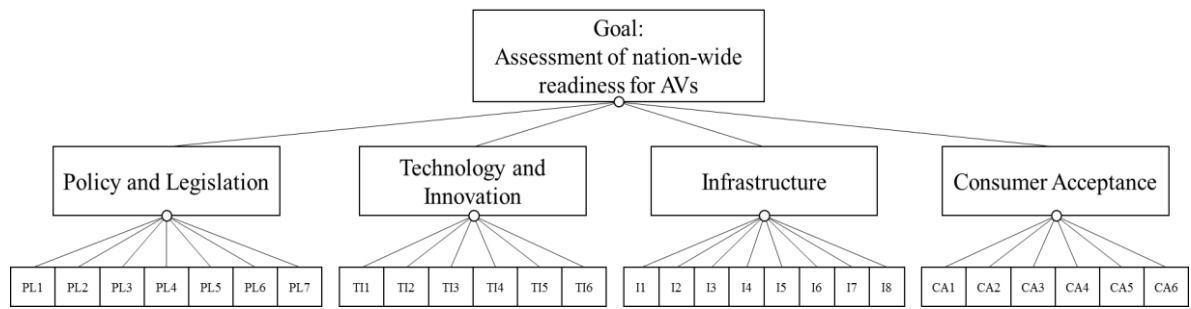

*Figure 2: Tiered criteria hierarchy*

*Table 4: Weighting of adopted criteria*

| Category | Criterion ID | Weight | Consensus | Aggregated weight | Rank |
|---|---|---|---|---|---|
| Policy and Legislation | | 0,236 | 78,5% | | |
| | PL1 | 0,102 | 86,4% | 0,024 | 18 |
| | PL2 | 0,055 | | 0,013 | 24 |
| | PL3 | 0,092 | | 0,022 | 20 |
| | PL4 | 0,194 | | 0,046 | 9 |
| | PL5 | 0,198 | | 0,047 | 8 |
| | PL6 | 0,228 | | 0,054 | 5 |
| | PL7 | 0,130 | | 0,031 | 14 |
| Technology and Innovation | | 0,408 | 78,5% | | |
| | TI1 | 0,137 | 92,9% | 0,056 | 4 |
| | TI2 | 0,184 | | 0,075 | 3 |
| | TI3 | 0,087 | | 0,035 | 13 |
| | TI4 | 0,299 | | 0,122 | 1 |
| | TI5 | 0,202 | | 0,082 | 2 |
| | TI6 | 0,091 | | 0,037 | 11 |
| Infrastructure | | 0,155 | 78,5% | | |
| | I1 | 0,109 | 91,8% | 0,017 | 21 |
| | I2 | 0,168 | | 0,026 | 17 |
| | I3 | 0,095 | | 0,015 | 23 |
| | I4 | 0,061 | | 0,009 | 27 |
| | I5 | 0,255 | | 0,040 | 10 |
| | I6 | 0,170 | | 0,026 | 16 |
| | I7 | 0,079 | | 0,012 | 25 |
| | I8 | 0,064 | | 0,010 | 26 |
| Consumer Acceptance | | 0,201 | 78,5% | | |
| | CA1 | 0,118 | 86,6% | 0,024 | 19 |
| | CA2 | 0,081 | | 0,016 | 22 |
| | CA3 | 0,247 | | 0,050 | 6 |
| | CA4 | 0,243 | | 0,049 | 7 |
| | CA5 | 0,132 | | 0,027 | 15 |
| | CA6 | 0,179 | | 0,036 | 12 |

It should be noted that the consensus rate is consistently at acceptable levels (above 75%). Moreover, it is noteworthy that for the category that was deemed by a large margin (40,2%) as the most important of the four, the consensus rate is near perfect (92,9%). The aggregated weights do not take into account that the various categories do not have the same number of criteria under them. However, this simplification is deemed as acceptable since the variance is low with a minimum number of criteria per category being six and the maximum eight and no category deviates significantly from the average number of criteria (6,75). Finally, it must be noted that the inconsistency ratio for all conducted AHP sessions is well within limits, namely under 10%.

The top three criteria from highest to lowest criticality are deemed to be the following:
- Industry investments in AV technologies
- Availability of the latest technology and capacity for innovation
- Number of non-governmental AV actors

On the other end of the spectrum, the bottom three criteria from lowest to highest criticality are deemed to be the following:
- Logistics infrastructure
- Clarity and level of standardization of road markings, signals and signage
- Shareability of construction plans

## 5. Assessing the case of Greece

This section applies the assessment methodology described in the previous sections. In order to promote objectivity, criteria for which no information is readily available are excluded from the analysis. Thus, the weights of the criteria will be normalized based on the weights of the excluded criteria in an effort to not unfairly promote or demote the readiness of Greece.

For each criterion a characterization of "very low", "low", "moderate", "high", and "very high" is applied. The scores that are associated respectively with each of these characterizations are shown in Table 5. It is obvious that the lowest possible score for the agglomeration of all criteria and categories after applying the weights is that of 0, while the highest achievable score is that of 1,00.

*Table 5: Scores based on criterion characterizations*

| Characterization | Score |
|---|---|
| Very low | 0 |
| Low | 0,25 |
| Moderate | 0,50 |
| High | 0,75 |
| Very high | 1,00 |

### 5.1 Policy and Legislation

Table 6 depicts the scores achieved by Greece for each of the criteria under the "Policy and Legislation" category. The criterion PL7 was excluded, since no official source was located adequate for the purposes of its assessment and the remaining weights were normalized accordingly. It may be observed that Greece achieves 42,7% of the maximum possible score in this category.

*Table 6: Assessment of Greece's readiness in regards to "policy and legislation"*

| ID | (Normalized) weight | Score | Weighted score | Justification |
|---|---|---|---|---|
| PL1 | 0,028 | 0,25 | 0,00693 | Only a single case, namely that regarding actions in the city of Trikala was located (L. 4313/2014). |
| PL2 | 0,015 | 0,25 | 0,00373 | Only a single pilot was identified, namely the one in the city of Trikala |
| PL3 | 0,025 | 0,25 | 0,00625 | Two entities exist within the Ministry of Transport and Infrastructure regarding ITS regulation and promotion (P.D. 123/2017) |
| PL4 | 0,053 | 0,75 | 0,03951 | This score is based on ranking of Greece (54 out of 136) in KPMG's change readiness index (2017) |
| PL5 | 0,054 | 0,25 | 0,01344 | This score was based on two rankings of 112 out 139 and 86 out of 139 on the "effectiveness of law-making bodies" and the "efficiency of legal framework in challenging regulations" respectively (WEF, 2018) |
| PL6 | 0,064 | 0,50 | 0,03096 | Currently, the National Access Point is under launch and it is expected to promote the concept of transport data sharing (Ayfadopolou and Mitsakis, 2018) |

*5.2 Technology and Innovation*

Table 7 depicts the scores achieved by Greece for each of the criteria under the "Technology and Innovation" category. Greece scores extremely low in this category achieving almost 12% of the maximum possible score.

*Table 7: Assessment of Greece's readiness in regards to "technology and innovation"*

| ID | (Normalized) weight | Score | Weighted score | Justification |
|---|---|---|---|---|
| TI1 | 0,056 | 0 | 0 | No such partnership was identified in the relevant literature (Silver, 2017) |

| ID | (Normalized) weight | Score | Weighted score | Justification |
|---|---|---|---|---|
| TI2 | 0,075 | 0,25 | 0,01877 | Currently, three research organizations are actively interested in AV technologies, namely CERTH-HIT, ICSS, and ITS Hellas |
| TI3 | 0,035 | 0,25 | 0,00887 | Only a single relevant patent has been identified in the "Google Patent" search engine using the "autonomous OR driverless OR self-driving vehicle" query |
| TI4 | 0,122 | 0 | 0 | According to the "Crunchbase" database, there are no headquarters of relevant commercial entities either located in Greece or belonging to Greek interests |
| TI5 | 0,082 | 0,25 | 0,02060 | This score was based on two rankings of 56 out 139 and 111 out of 139 on the "availability of latest technologies" and the "capacity for innovation" respectively (WEF, 2018) |
| TI6 | 0,037 | 0 | 0 | This was deemed as extremely low since according the relevant literature (EAMA, 2017; EAFO, 2019) the market share is significantly lower than 10% |

## 5.3 Infrastructure

Table 8 depicts the scores achieved by Greece for each of the criteria under the "Infrastructure" category. The criteria I6, I7, and I8 were excluded, since no official sources were located adequate for the purposes of its assessment and the remaining weights were normalized accordingly. It may be observed that Greece achieves almost 52% of the maximum possible score in this category.

*Table 8: Assessment of Greece's readiness in regards to "infrastructure"*

| ID | (Normalized) weight | Score | Weighted score | Justification |
|---|---|---|---|---|
| I1 | 0,025 | 0,25 | 0,00614 | This score is based on the low density of EV charging stations (according to "charging.gr" database) as well as on the limited regulated parking places |
| I2 | 0,038 | 0,50 | 0,01892 | This score is based on the relatively high ranking of Greece (37 out of 88) on a 4G coverage assessment (OpenSignal, 2018) |

| ID | (Normalized) weight | Score | Weighted score | Justification |
|---|---|---|---|---|
| I3 | 0,021 | 0,75 | 0,01605 | This score is based on a ranking of 36 out of 140 on an assessment with the same name conducted by WEF (2018) |
| I4 | 0,014 | 0,75 | 0,01031 | This score is based on a ranking of 42 out of 160 on a relevant assessment conducted by World Bank Group (2018) |
| I5 | 0,057 | 0,50 | 0,02872 | This score is based on a ranking of 65 out of 140 on a change readiness assessment conducted by KPMG (2019) |

## 5.4 Consumer Acceptance

Table 9 depicts the scores achieved by Greece for each of the criteria under the "Consumer Acceptance" category. It may be observed that Greece achieves 46,4% of the maximum possible score in this category.

*Table 9:* *Assessment of Greece's readiness in regards to "consumer acceptance"*

| ID | (Normalized) weight | Score | Weighted score | Justification |
|---|---|---|---|---|
| CA1 | 0,024 | 0,50 | 0,01186 | This score is based on a survey conducted by Dolianitis et al. (2019); while the acceptance in this paper appears as high, the score is characterized as moderate since the sample was limited and restricted to interviewees of high educational levels |
| CA2 | 0,016 | 0,25 | 0,00407 | Only the population of the city of Trikala falls under this characterization |
| CA3 | 0,050 | 0,75 | 0,03724 | This score is based on a ranking of 39 out of 140 on a change readiness assessment conducted by KPMG (2019) |
| CA4 | 0,049 | 0,50 | 0,02442 | This score is based on a ranking of 57 out of 140 on a relevant assessment conducted by WEF (2018) |
| CA5 | 0,027 | 0,25 | 0,00663 | This score is based on a report concerning ride-hailing market revenue by Statista (2019) |
| CA6 | 0,036 | 0,25 | 0,00899 | This score is based on a report concerning average prices of EU car sales by Statista (2017) |

## 5.5 Readiness index of Greece

The overall readiness index of Greece appears to be equal to 32%. While this score corresponds to a "low" readiness level, it must be noted that scores fairly well in three of the categories in which it is assessed. However, Greece scores particularly low in what was deemed as the most important category, namely that of "Technology and Innovation". Figure 3 depicts the rating Greece achieved both as a whole and in individual categories.

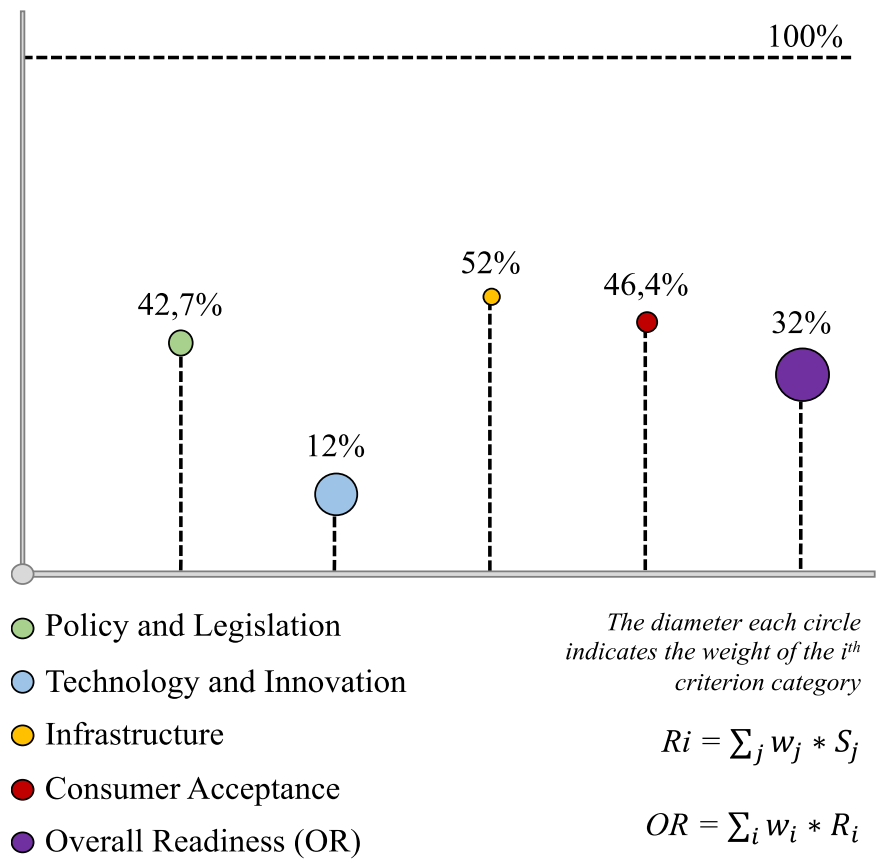

*Figure 3: Readiness for AV technologies – the case of Greece*

## 6. Conclusions

This paper provides interested parties with a framework on how to assess the readiness of countries for adopting AV technologies. An attempt was made to combine and enrich existing methodologies in structured and transparent way. A tiered AHP methodology was adopted in an effort to further contribute to existing methodologies by deriving weighting factors for each criterion and criterion category.

The results concerning the case of Greece while at firstly glance do not appear optimistic, it is of note that Greece performed disappointingly only in one category. This presents a clear area for policy makers to focus on, namely promoting the creation of companies that are engaged in this field, promoting partnerships with existing foreign companies, and finally promoting the acquisition of funds for investment in this field.

Future research will update this assessment and include criteria for which no data were currently found. Furthermore, future research will use this methodology to benchmark Greece against both countries leading the field (e.g. The Netherlands) as well as countries comparable to Greece. Finally, since the adopted methodology relies upon the input of a group of experts the inclusion of additional interviewees of additional disciplines and backgrounds would allow for results that are more focused to particular audiencies (e.g. policy makers, manufacturers). This would also allow for the analysis to be focused on specific contexts, such as having an urban or rural environment distinction.

## *7. References-Bibliography*

Ayfadopoulou, G., & Mitsakis, E. (2018). National Access Points - Current status at EU level and progress of the NAP in Greece. Retrieved from: https://www.its-hellas.gr/images/PDF/DiimeridaDEC2018/Presentations/Day_1/2._Best_practices/6._NAP_presentation_v9.pdf

Dolianitis, A., Chalkiadakis, C., Mylonas, C., & Tzanis, D. (2019). How Will Autonomous Vehicles Operate in an Unlawful Environment? - The Potential of Autonomous Vehicles for Disregarding the Law. Presented at 6[th] International Conference on Models and Technologies for Intelligent Transportation Systems(MT-ITS 2019).

European Alternative Fuels Observatory (EAFO) (2019). Greece. Retrieved from: https://www.eafo.eu/countries/greece/1735/summary

European Automobile Manufacturers Association (EAMA) (2017). ACEA Report Vehicles in use Europe 2017. Retrieved from: https://www.acea.be/uploads/statistic_documents/ACEA_Report_Vehicles_in_use-Europe_2017.pdf

Fagnant, D. J., & Kockelman, K. (2015). Preparing a nation for autonomous vehicles: Opportunities, barriers and policy recommendations. Transportation Research Part A: Policy and Practice, 77, 167–181. https://doi.org/10.1016/j.tra.2015.04.003

Goepel, K. D. (2018). New AHP Excel template with multiple inputs. Retrieved from: https://bpmsg.com/new-ahp-excel-template-with-multiple-inputs/

Henderson, J. (2018). Assessing the Readiness of Autonomous Vehicles. Retrieved from: https://www.vprobroadcast.com/.../Jim%20Henderson-%...

Johnson, C. (2017). Readiness of the road network for connected and autonomous vehicles. Retrieved from: https://www.racfoundation.org/wp-content/uploads/2017/11/CAS_Readiness_of_the_road_network_April_2017.pdf

Kimley-Horn (2016). NC readiness for connected and autonomous vehicles (CAV) - Final Report. Retrieved from: https://transportationops.org/sites/transops/files/NC-Roadmap-for-CAV_Final_ALL.pdf

KPMG (2017). 2017 Change Readiness Index tool. Retrieved from: https://home.kpmg/xx/en/home/insights/2017/06/change-readiness-tool.html?countryCode=GR

KPMG (2019). 2019 Autonomous Vehicles Readiness Index - Assessing countries' preparedness for autonomous vehicles. Retrieved from: https://assets.kpmg/content/dam/kpmg/xx/pdf/2019/02/2019-autonomous-vehicles-readiness-index.pdf

Nunes, A., & Hernandez, K. (2019). The Cost of Self-Driving Cars Will Be the Biggest Barrier to Their Adoption. Retrieved from: https://hbr.org/2019/01/the-cost-of-self-driving-cars-will-be-the-biggest-barrier-to-their-adoption

OpenSignal (2018). The State of LTE (February 2018). Retrieved from: https://www.opensignal.com/reports/2018/02/state-of-lte

Rodrigue, J. P. (2017). The Geography of Transport Systems. Fourth Edition. New York: Routledge.

Saaty R.W. (1987). the Analytic Hierarchy Process-What and How It Is Used It Is. International Journal of Advanced Science and Technology, 9(1), 19–24. https://doi.org/https://doi.org/10.1016/0270-0255(87)90473-8

Statista (2017). Average price (including tax) of passenger cars in the EU in 2013 and 2017, by country* (in euros). Retrieved from: https://www.statista.com/statistics/425095/eu-car-sales-average-prices-in-by-country/

Statista (2019). Ride Hailing - Greece. Retrieved from: https://www.statista.com/outlook/368/138/ride-hailing/greece

World Bank Group (2018). Aggregated LPI 2012-2018. Retrieved from: https://lpi.worldbank.org/international/aggregated-ranking

World Economic Forum (WEF) (2018). The Global Competitiveness Report 2018. Retrieved from: http://www3.weforum.org/docs/GCR2018/05FullReport/TheGlobalCompetitivenessReport2018.pdf